\documentclass[12pt,titlepage]{article}
\usepackage{amsmath}
\usepackage{amssymb}
\usepackage[dvips]{graphicx}
\usepackage{amsfonts}
\usepackage[portuguese]{babel}
\usepackage[font=small,format=plain,labelfont=bf,up,textfont=it,up]{caption}

\begin{document}

\title{\textbf{Uma breve discuss\~{a}o sobre os poss\'{\i}veis estados
ligados para uma classe de potenciais singulares}\thanks{%
To appear in Revista Brasileira de Ensino de F\'{\i}sica}\\
{\small (A brief discussion on the possible bound states for a class of
singular potentials)}}
\date{}
\author{Douglas R.M. Pimentel\thanks{%
E-mail: douglas.roberto.fis@gmail.com.} e Antonio S. de Castro\thanks{%
E-mail: castro@pq.cnpq.br.} \\
\\
Departamento de F\'{\i}sica e Qu\'{\i}mica, \\
Universidade Estadual Paulista \textquotedblleft J\'{u}lio de Mesquita
Filho\textquotedblright, \\
Guaratinguet\'{a}, SP, Brasil}
\maketitle

\begin{abstract}
Investiga-se a equa\c{c}\~{a}o de Schr\"{o}dinger unidimensional com uma
classe de potenciais $V\left( |x|\right) $ que se anulam no infinito e
apresentam singularidade dominante na origem na forma $\alpha /|x|^{\beta }$
($0<\beta \leq 2$). A hermiticidade dos operadores associados com
quantidades f\'{\i}sicas observ\'{a}veis \'{e} usada para determinar as condi%
\c{c}\~{o}es de contorno apropriadas. Dupla degeneresc\^{e}ncia e exclus\~{a}%
o de solu\c{c}\~{o}es sim\'{e}tricas, consoante o valor de $\beta $, s\~{a}o
discutidas. Solu\c{c}\~{o}es expl\'{\i}citas para o \'{a}tomo de hidrog\^{e}%
nio e o potencial de Kratzer s\~{a}o apresentadas.

\bigskip

\noindent \textbf{Palavras-chave:} potencial singular, degeneresc\^{e}ncia,
\'{a}tomo de hidrog\^{e}nio unidimensional, potencial de Kratzer, colapso
para o centro \newline

\bigskip

\bigskip

\bigskip

\noindent {\small {The one-dimensional Schr\"{o}dinger equation for a class
of potentials }}$V(|x|)${\small {\ which vanish at infinity and present
dominant singularity at the origin in the form }}$\alpha /|x|^{\beta }$ ($%
0<\beta \leq 2$) {\small {is investigated. The Hermiticity of the operators
related to observable physical quantities is used to determinate the proper
boundary conditions. Double degeneracy and exclusion of symmetric solutions,
consonant the value of }}$\beta $, are discussed. Explicit solutions for the
hydrogen atom and the Kratzer potential are presented.

\bigskip

\end{abstract}

\section{Introdu\c{c}\~{a}o}

O problema geral de espalhamento e estados ligados em potenciais singulares
\'{e} um tema antigo e recorrente em mec\^{a}nica qu\^{a}ntica (veja, e.g.,
\cite{cas}). Recentemente, o oscilador harm\^{o}nico singular foi esmiu\c{c}%
ado nesta Revista \cite{rbef}, e revelou-se uma opul\^{e}ncia de conceitos e
t\'{e}cnicas que s\~{a}o da maior import\^{a}ncia para os estudantes e
instrutores de mec\^{a}nica qu\^{a}ntica e f\'{\i}sica matem\'{a}tica.

O problema com o potencial $V\left( |x|\right) =-|\alpha |/|x|$, conhecido
como \'{a}tomo de hidrog\^{e}nio unidimensional, tem recebido consider\'{a}%
vel aten\c{c}\~{a}o na literatura por mais de cinquenta anos (para uma ampla
lista de refer\^{e}ncias relacionadas com celeumas e aplica\c{c}\~{o}es em f%
\'{\i}sica at\^{o}mica, f\'{\i}sica molecular e em f\'{\i}sica da mat\'{e}%
ria condensada veja, e.g., \cite{xia} e \cite{yep}). O problema tamb\'{e}m
tem sido objeto de investiga\c{c}\~{a}o em diversos contextos relativ\'{\i}%
sticos \cite{spe}-\cite{bar}. O potencial da forma $\alpha _{1}/|x|+\alpha
_{2}/x^{2}$, com $x\in \lbrack 0,\infty )$ e $\alpha _{1}<0$ e $\alpha
_{2}>0 $, conhecido como potencial de Kratzer, tem sido usado na descri\c{c}%
\~{a}o do espectro molecular \cite{kra}, e tamb\'{e}m no estudo de transfer%
\^{e}ncia de energia vibracional de mol\'{e}culas poliat\^{o}micas \cite%
{ulrich}. O problema de estados ligados da equa\c{c}\~{a}o de Schr\"{o}%
dinger com o potencial de Kratzer \'{e} um problema analiticamente sol\'{u}%
vel, e sua solu\c{c}\~{a}o pode ser encontrada em livros-texto (\cite{bg} e
\cite{lan}, por exemplo). O caso unidimensional, com par\^{a}metros $\alpha
_{1}$ e $\alpha _{2}\ $arbitr\'{a}rios, tamb\'{e}m \'{e} um problema
analiticamente sol\'{u}vel, e tem sido investigado no \^{a}mbito de equa\c{c}%
\~{o}es relativ\'{\i}sticas \cite{asc1}, \cite{asc2}.

Na esteira pedag\'{o}gica da Ref. \cite{rbef}, abordamos neste trabalho o
problema de estados ligados no \^{a}mbito da equa\c{c}\~{a}o de Schr\"{o}%
dinger unidimensional para uma classe de potenciais $V\left( |x|\right) $
que se anulam no infinito e apresentam singularidade dominante na origem na
forma $\alpha /|x|^{\beta }$ ($0<\beta \leq 2$). Visto que $x=0$ \'{e} um
ponto singular da equa\c{c}\~{a}o de Schr\"{o}dinger, a determina\c{c}\~{a}o
de suas solu\c{c}\~{o}es requer uma an\'{a}lise cuidadosa na vi\-zi\-nhan%
\c{c}a da origem. A hermiticidade dos operadores associados com quantidades f%
\'{\i}sicas observ\'{a}veis p\~{o}e \`{a} mostra as condi\c{c}\~{o}es de
contorno apropriadas. Demonstramos que o espectro do problema definido no
semieixo \'{e} n\~{a}o-degenerado e que n\~{a}o h\'{a} estados ligados se o
potencial com comportamento dominante na origem do tipo inversamente quadr%
\'{a}tico for fortemente atrativo. A extens\~{a}o do problema para todo o
eixo revela um espectro n\~{a}o-degenerado se $\beta <2$, e degenerado se $%
\beta =2$. Revela-se tamb\'{e}m que somente autofun\c{c}\~{o}es antissim\'{e}%
tricas ocorrem no caso $1\leq \beta <2$, e que autofun\c{c}\~{o}es sim\'{e}%
tricas e antissim\'{e}tricas ocorrem no caso $\beta <1$. A seguir
investigamos as solu\c{c}\~{o}es para $V\left( |x|\right) $ com formas bem
definidas. Come\c{c}amos com o problema definido no semieixo, e depois de
fatorar o comportamento da autofun\c{c}\~{a}o na origem e no infinito, a equa%
\c{c}\~{a}o de Schr\"{o}dinger se transmuta em uma equa\c{c}\~{a}o
diferencial hipergeom\'{e}trica confluente para a classe de potencias de
interesse. Demonstramos que um conjunto infinito de estados ligados tem
presen\c{c}a no caso do potencial singular atrativo do tipo $1/|x|$, ainda
que o potencial seja fortemente atrativo. Tamb\'{e}m mostramos que no caso
da adi\c{c}\~{a}o de um termo singular fracamente atrativo do tipo $1/x^{2}$
n\~{a}o h\'{a} cabimento em se falar em colapso para a origem (como afirmado
em \cite{bg}) ou inexist\^{e}ncia de estado fundamental, e que h\'{a} um
conjunto infinito de estados ligados. Dentro da classe de potenciais
investigados neste trabalho, demonstramos a inexist\^{e}ncia de estados
ligados com o potencial $V\left( |x|\right) =\alpha /x^{2}$, e apresentamos
a solu\c{c}\~{a}o expl\'{\i}cita dos estados ligados com o potencial $%
V\left( |x|\right) =\alpha _{1}/|x|+\alpha _{2}/x^{2}$, com $\alpha _{1}\neq
0$.

\section{Potenciais singulares na origem}

A equa\c{c}\~{a}o de Schr\"{o}dinger unidimensional independente do tempo
para uma part\'{\i}cula de massa $m$ sujeita a um potencial externo $V\left(
x\right) $ \'{e} dada por%
\begin{equation}
H\psi =E\psi ,  \label{auto}
\end{equation}%
onde $H$ \'{e} o operador hamiltoniano
\begin{equation}
H=-\frac{\hbar ^{2}}{2m}\frac{d^{2}}{dx^{2}}+V.  \label{1}
\end{equation}%
Aqui, $\psi \left( x\right) $ \'{e} a autofun\c{c}\~{a}o que descreve o
estado estacion\'{a}rio, $E$ \'{e} a autoenergia, e $\hbar $ \'{e} a
constante de Planck reduzida ($\hbar =h/(2\pi )$). A equa\c{c}\~{a}o de Schr%
\"{o}dinger tamb\'{e}m pode ser escrita na forma
\begin{equation}
\frac{d^{2}\psi }{dx^{2}}-\left( k^{2}+\frac{2mV}{\hbar ^{2}}\right) \psi =0,
\label{3}
\end{equation}%
\noindent com%
\begin{equation}
k^{2}=-\frac{2mE}{\hbar ^{2}}.  \label{k2}
\end{equation}%
A quantidade%
\begin{equation}
\rho =\left\vert \psi \right\vert ^{2}\quad  \label{rho}
\end{equation}%
\'{e} interpretada como sendo a densidade de probabilidade. Com condi\c{c}%
\~{o}es de contorno apropriadas, o problema se reduz \`{a} determina\c{c}%
\~{a}o do par caracter\'{\i}stico $\left( E,\psi \right) $ de uma equa\c{c}%
\~{a}o do tipo Sturm-Liouville (veja, e.g., \cite{but}). Todas as
quantidades f\'{\i}sicas observ\'{a}veis correspondem a operadores
hermitianos. Para autofun\c{c}\~{o}es definidas no intervalo $[x_{1},x_{2}]$%
, o operador $\mathcal{O}$ \'{e} dito ser hermitiano se%
\begin{equation}
\int_{x_{1}}^{x_{2}}dx\,\left( \mathcal{O}\psi _{1}\right) ^{\ast }\psi
_{2}=\int_{x_{1}}^{x_{2}}dx\,\psi _{1}^{\ast }\left( \mathcal{O}\psi
_{2}\right) ,
\end{equation}%
onde $\psi _{1}$ e $\psi _{2}$ s\~{a}o duas autofun\c{c}\~{o}es quaisquer
que fazem $\int_{x_{1}}^{x_{2}}dx\,\psi _{1}^{\ast }\left( \mathcal{O}\psi
_{2}\right) <\infty $. Em particular, as autofun\c{c}\~{o}es devem ser
quadrado-integr\'{a}veis, viz. $\int_{x_{1}}^{x_{2}}dx\,|\psi |^{2}$ $%
<\infty $.

Neste trabalho focalizamos nossa aten\c{c}\~{a}o nos estados ligados com
potenciais que t\^{e}m o comportamento%
\begin{equation}
V\left( |x|\right) \sim \left\{
\begin{array}{c}
\frac{\alpha }{|x|^{\beta }}, \\
\\
0,%
\end{array}%
\begin{array}{c}
|x|\rightarrow 0 \\
\\
|x|\rightarrow \infty ,%
\end{array}%
\right.  \label{pot}
\end{equation}%
com $0<\beta \leq 2$, e consideramos autofun\c{c}\~{o}es definidas no
intervalo $[0,\infty )$. Neste caso, a equa\c{c}\~{a}o de Schr\"{o}dinger
\'{e} uma equa\c{c}\~{a}o diferencial singular e a autofun\c{c}\~{a}o pode
manifestar algum comportamento patol\'{o}gico. Tal sequela poderia
comprometer a exist\^{e}ncia de integrais do tipo $\int_{0}^{\infty
}dx\,\psi _{1}^{\ast }\left( \mathcal{O}\psi _{2}\right) $, e assim
comprometer a hermiticidade dos operadores associados com as quantidades f%
\'{\i}sicas observ\'{a}veis. Por isto, o comportamento das solu\c{c}\~{o}es
de (\ref{3}) na vizinhan\c{c}a da origem exige muita aten\c{c}\~{a}o. Visto
que os estados ligados constituem uma classe de solu\c{c}\~{o}es da equa\c{c}%
\~{a}o de Schr\"{o}dinger que representam um sistema localizado numa regi%
\~{a}o finita do espa\c{c}o, devemos procurar autofun\c{c}\~{o}es que se
anulam \`{a} medida que $|x|\rightarrow \infty $. Tamb\'{e}m, neste caso
podemos normalizar $\psi $ fazendo $\int_{0}^{\infty }dx\,|\psi |^{2}=1$.

Na vizinhan\c{c}a da origem a equa\c{c}\~{a}o (\ref{3}) passa a ter a forma

\begin{equation}
\frac{d^{2}\psi }{dx^{2}}-\frac{2m\alpha }{\hbar ^{2}|x|^{\beta }}\,\psi
\simeq 0,  \label{SOL}
\end{equation}%
e no semieixo positivo podemos escrever a solu\c{c}\~{a}o geral de (\ref{SOL}%
) como%
\begin{equation}
\psi \simeq \left\{
\begin{array}{c}
A\,|x|^{s_{+}}+B\,|x|^{s_{-}},\quad \text{para }s_{+}\neq s_{-} \\
\\
C\,\,|x|^{1/2}+D\,\,|x|^{1/2}\log |x|,\quad \text{para }s_{+}=s_{-},%
\end{array}%
\right.  \label{original}
\end{equation}%
onde $A$, $B$, $C$ e $D$ s\~{a}o constantes arbitr\'{a}rias, e $s_{\pm }$
\'{e} solu\c{c}\~{a}o da equa\c{c}\~{a}o indicial%
\begin{equation}
s_{\pm }\left( s_{\pm }-1\right) -\frac{2m\alpha }{\hbar ^{2}|x|^{\beta -2}}%
\simeq 0.  \label{indi}
\end{equation}%
A equa\c{c}\~{a}o indicial resulta da considera\c{c}\~{a}o dos termos de
mais baixa ordem da expans\~{a}o em s\'{e}ries de pot\^{e}ncias. Note que
esta equa\c{c}\~{a}o faz sentido somente se $\beta \leq 2$, i.e., se a
origem for uma singularidade n\~{a}o-essencial.\footnote{%
A origem \'{e} uma singularidade n\~{a}o-essencial (ou regular) se $%
x^{2}V\left( |x|\right) $ for finita no limite $|x|\rightarrow 0$. Caso contr%
\'{a}rio, a origem \'{e} uma singularidade essencial (ou irregular). No caso
de singularidade n\~{a}o-essencial, o teorema de Fuchs (veja, e.g. \cite{but}%
) garante que a solu\c{c}\~{a}o geral da equa\c{c}\~{a}o diferencial \'{e} a
combina\c{c}\~{a}o linear de duas solu\c{c}\~{o}es linearmente independentes
$c_{1}S_{1}\left( x\right) +c_{2}S_{2}\left( x\right) $, onde $S_{1}\left(
x\right) $ e $S_{2}\left( x\right) $ s\~{a}o express\'{\i}veis como s\'{e}%
ries de pot\^{e}ncias em torno da origem, ou $c_{1}S_{1}\left( x\right)
+c_{2}\left( S_{1}\left( x\right) \log |x|+S_{2}\left( x\right) \right) $.}
Assim sendo, os estados ligados em potenciais com singularidade mais forte
que $1/x^{2}$ est\~{a}o exclu\'{\i}dos de nossa considera\c{c}\~{a}o. A equa%
\c{c}\~{a}o indicial tem as solu\c{c}\~{o}es%
\begin{equation}
s_{\pm }=\frac{1}{2}\left( 1\pm \sqrt{1+\frac{8m\alpha }{\hbar ^{2}}}\right)
,\quad \text{para }\beta =2\text{,}  \label{rad}
\end{equation}%
e%
\begin{equation}
s_{+}=1\text{ e }s_{-}=0,\quad \text{para }\beta <2.  \label{rad1}
\end{equation}%
Na vizinhan\c{c}a da origem, o comportamento do termo
\begin{equation}
V_{\tilde{n}n}=\psi _{\tilde{n}}^{\ast }\frac{\alpha }{|x|^{\beta }}\psi _{n}
\notag
\end{equation}%
amea\c{c}a a hermiticidade do operador associado com a energia potencial.
Para $\beta =2$ e $s_{+}\neq s_{-}$, podemos escrever%
\begin{eqnarray}
V_{\tilde{n}n} &\simeq &\frac{\alpha }{|x|^{\beta }}\left[ A_{\tilde{n}%
}^{\ast }A_{n}\,|x|^{2\text{{\small Re}}s_{+}}+B_{\tilde{n}}^{\ast
}B_{n}\,|x|^{2\text{{\small Re}}s_{-}}\right.  \notag \\
&& \\
&&+\left. A_{\tilde{n}}^{\ast }B_{n}\,|x|^{s_{+}^{\ast }+s_{-}}+B_{\tilde{n}%
}^{\ast }A_{n}\,|x|^{s_{+}+s_{-}^{\ast }}\right] ,  \notag
\end{eqnarray}%
e para $\beta =2$ e $s_{+}=s_{-}$%
\begin{eqnarray}
V_{\tilde{n}n} &\simeq &\frac{\alpha }{|x|^{\beta }}\,\left[ C_{\tilde{n}%
}^{\ast }C_{n}+D_{\tilde{n}}^{\ast }D_{n}\log ^{2}|x|\right.  \notag \\
&& \\
&&+\left. \left( C_{\tilde{n}}^{\ast }D_{n}+D_{\tilde{n}}^{\ast
}C_{n}\right) \log |x|\right] .  \notag
\end{eqnarray}%
Vemos destas \'{u}ltimas rela\c{c}\~{o}es que a hermiticidade do operador
associado com a energia potencial \'{e} verificada somente se \textrm{%
Re\thinspace }$s_{\pm }>1/2$, o que equivale a dizer que o sinal negativo
defronte do radical em (\ref{rad}) deve ser descartado e $\alpha $ deve ser
maior que $\alpha _{c}$, com%
\begin{equation}
\alpha _{c}=-\frac{\hbar ^{2}}{8m}.
\end{equation}%
Note que a solu\c{c}\~{a}o expressa pela segunda linha de (\ref{original}),
correspondente \`{a} raiz dupla da equa\c{c}\~{a}o indicial, perde sua
serventia. Para $\beta <2$ temos que $V_{\tilde{n}n}$ \'{e} integr\'{a}vel
somente se $\psi $ na primeira linha de (\ref{original}) tiver $B=0$ para $%
1\leq \beta <2$. Porque a equa\c{c}\~{a}o de Schr\"{o}dinger \'{e} linear e
sua solu\c{c}\~{a}o pode envolver apenas uma constante de integra\c{c}\~{a}o
multiplicativa, a ser determinada por interm\'{e}dio da condi\c{c}\~{a}o de
normaliza\c{c}\~{a}o, devemos ter $A=0$ ou $B=0$ para $\beta <1$.

Diante do exposto, podemos afirmar que $\psi $ se comporta na vizinhan\c{c}a
da origem como%
\begin{equation}
F|x|^{s},
\end{equation}%
onde $F$ \'{e} uma constante arbitr\'{a}ria, e $s$ \'{e} uma quantidade real
com os valores poss\'{\i}veis segregados como segue:
\begin{equation}
s=\left\{
\begin{array}{c}
\frac{1}{2}\left( 1+\sqrt{1+\frac{8m\alpha }{\hbar ^{2}}}\right) , \\
\\
1, \\
\\
0\text{ ou }1,%
\end{array}%
\begin{array}{c}
\quad \text{para }\beta =2\text{ e }\alpha >\alpha _{c} \\
\\
\text{para }1\leq \beta <2 \\
\\
\text{para }\beta <1.%
\end{array}%
\right. \text{ }
\end{equation}%
O crit\'{e}rio de hermiticidade do operador associado com a energia
potencial \'{e} l\'{\i}cito e suficiente para descartar solu\c{c}\~{o}es esp%
\'{u}rias. A ortonormalizabilidade das autofun\c{c}\~{o}es (relacionada com
a hermiticidade do operador hamiltoniano) e a hermiticidade do o\-pe\-rador
momento s\~{a}o crit\'{e}rios mais fr\'{a}geis porque envolvem o
comportamento de $\psi _{\tilde{n}}^{\ast }\psi _{n}$ e $\psi _{\tilde{n}%
}^{\ast }d\psi _{n}/dx$ na vizinhan\c{c}a da origem, respectivamente. O
requerimento da hermiticidade do operador energia cin\'{e}tica \'{e} t\~{a}o
r\'{\i}gido quanto o requerimento da hermiticidade do operador energia
potencial para $\beta =2$, por\'{e}m \'{e} mais fr\'{a}gil para $\beta <2$.
A bem da verdade, a hermiticidade do hamiltoniano tem sido usada com sucesso
na descri\c{c}\~{a}o do espectro do \'{a}tomo de hidrog\^{e}nio
unidimensional \cite{xia} tanto quanto para desmascarar o hidrino (estado
estramb\'{o}tico do hidrog\^{e}nio com energia mais baixa que aquela de seu
estado fundamental normal) \cite{dom}. \'{E} instrutivo notar que o
comportamento da autofun\c{c}\~{a}o na vizinhan\c{c}a da origem independe da
intensidade do potencial, caracterizada pelo par\^{a}metro $\alpha $, no
caso em que o potencial \'{e} menos singular que $1/x^{2}$, e que a condi%
\c{c}\~{a}o de Dirichlet homog\^{e}nea ($\psi \left( 0_{+}\right) =0$) \'{e}
essencial sempre que $1\leq \beta \leq 2$ ($s>1/2$), contudo ela tamb\'{e}m
ocorre para $\beta <1$ quando $s=1$ mas n\~{a}o para $s=0$. Estes resultados
est\~{a}o sumarizados na Tabela 1.

\begin{table}[ht]
\begin{center}
\begin{tabular}{|c||c|c|}
\hline
&  &  \\
& $\left. \psi \right\vert _{x=0_{+}}$ & $\left. \frac{d\psi }{dx}%
\right\vert _{x=0_{+}}$ \\
&  &  \\ \hline\hline
&  &  \\
$\beta =2\text{ e }\alpha >0$ & $0$ & $0$ \\
&  &  \\ \hline
&  &  \\
$\beta =2\text{ e }\alpha <0$ & $0$ & $\infty $ \\
&  &  \\ \hline
&  &  \\
$1\leq \beta <2$ & $0$ & $<\infty $ \\
&  &  \\ \hline
&  &  \\
$\beta <1\text{ e }s=0$ & $<\infty $ & $0$ \\
&  &  \\ \hline
&  &  \\
$\beta <1\text{ e }s=1$ & $0$ & $<\infty $ \\
&  &  \\ \hline
\end{tabular}%
\end{center}
\caption{Autofun\c{c}\~{a}o e sua primeira derivada na vizinhan\c{c}a da
origem em fun\c{c}\~{a}o dos par\^{a}metros do potencial.}
\end{table}

A equa\c{c}\~{a}o de Schr\"{o}dinger independente do tempo para o nosso
problema, Eq. (\ref{3}), tem o comportamento para grandes valores de $|x|$
dado por $d^{2}\psi /dx^{2}-k^{2}\psi \simeq 0,$ e da\'{\i} sucede que a
forma assint\'{o}tica da solu\c{c}\~{a}o quadrado-integr\'{a}vel \'{e} dada
por
\begin{equation}
\psi \simeq e^{-k|x|},\quad k\in
\mathbb{R}
,  \label{asym}
\end{equation}%
onde $k$ est\'{a} definido por (\ref{k2}). Portanto, podemos afirmar que os
poss\'{\i}veis estados ligados t\^{e}m espectro negativo.

\section{Solu\c{c}\~{a}o no semieixo}

As condi\c{c}\~{o}es de contorno impostas sobre a autofun\c{c}\~{a}o nos
extremos do intervalo nos permite tirar conclus\~{o}es acerca da degeneresc%
\^{e}ncia. Com esta finalidade, seguimos os passos da Ref. \cite{lan}. Sejam
$\psi _{1}$ e $\psi _{2}$ duas autofun\c{c}\~{o}es correspondentes \`{a}
mesma energia. A equa\c{c}\~{a}o de Schr\"{o}dinger (\ref{3}) implica que%
\begin{equation}
\psi _{1}\frac{d^{2}\psi _{2}}{dx^{2}}-\psi _{2}\frac{d^{2}\psi _{1}}{dx^{2}}%
=0.  \label{w1}
\end{equation}%
Nesta circunst\^{a}ncia, a integral de (\ref{w1}) resulta em%
\begin{equation}
\psi _{1}\frac{d\psi _{2}}{dx}-\psi _{2}\frac{d\psi _{1}}{dx}=W\left( \psi
_{1},\psi _{2}\right) =W_{0}=\text{constante,}  \label{w2}
\end{equation}%
onde $W\left( \psi _{1},\psi _{2}\right) $ designa o wronskiano de $\psi
_{1} $ e $\psi _{2}$. O comportamento assint\'{o}tico de $\psi _{1}$ e $\psi
_{2}$ faz $W_{0}=0$, e assim o wronskiano \'{e} nulo em todo o semieixo,
tendo como consequ\^{e}ncia imediata a depend\^{e}ncia linear entre $\psi
_{1}$ e $\psi _{2}$. Conclui-se, ent\~{a}o, que o espectro \'{e} n\~{a}%
o-degenerado. No entanto, tal conclus\~{a}o \'{e} fidedigna somente se (\ref%
{w1}) for integr\'{a}vel. Na vizinhan\c{c}a da origem, cada parcela do lado
esquerdo de (\ref{w1}) \'{e} proporcional a $s\left( s-1\right) |x|^{2s-2}$,
sendo integr\'{a}vel somente se $s=0$ ou $s>1/2$. Destarte, a ado\c{c}\~{a}o
do bem-afortunado crit\'{e}rio de hermiticidade do operador associado com a
energia potencial garante a inexist\^{e}ncia de degeneresc\^{e}ncia ainda
que os potenciais sejam singulares.

Note que para o problema definido no semieixo mas limitado por uma barreira
infinita no semieixo complementar, a solu\c{c}\~{a}o com $s=0$ deve ser
banida por causa da continuidade da autofun\c{c}\~{a}o em $x=0$.

\section{Solu\c{c}\~{a}o em todo o eixo}

Com potenciais pares sob a troca de $x$ por $-x$, as autofun\c{c}\~{o}es em
todo o eixo, com paridades bem definidas, podem ser obtidas das autofun\c{c}%
\~{o}es definidas no semieixo por meio de extens\~{o}es sim\'{e}tricas e
antissim\'{e}tricas. A autofun\c{c}\~{a}o definida para todo o eixo $X$ pode
ser escrita como%
\begin{equation}
\psi ^{\left( p\right) }\left( x\right) =[\theta \left( x\right) +p\,\theta
\left( -x\right) ]\psi \left( |x|\right) ,  \label{todoeixo}
\end{equation}%
onde $p=\pm 1$ e $\theta \left( x\right) $ \'{e} a fun\c{c}\~{a}o degrau de
Heaviside ($1$ para $x>0$, e $0$ para $x<0$). Estas duas autofun\c{c}\~{o}es
linearmente independentes possuem a mesma energia, ent\~{a}o, em princ\'{\i}%
pio, existe uma dupla degeneresc\^{e}ncia. Observe que, apesar da nulidade
do wronskiano de $\psi ^{\left( +\right) }$ e $\psi ^{\left( -\right) }$,
como pode ser inferido pelo comportamento assint\'{o}tico, est\'{a} claro
que $\psi ^{\left( +\right) }$ n\~{a}o pode ser expressa em termos de $\psi
^{\left( -\right) }$, e vice-versa. Entretanto, temos de considerar as condi%
\c{c}\~{o}es de conex\~{a}o entre a autofun\c{c}\~{a}o, e tamb\'{e}m sua
derivada primeira, \`{a} direita e \`{a} esquerda da origem.

A autofun\c{c}\~{a}o deve ser cont\'{\i}nua na origem, do contr\'{a}rio o
valor esperado da energia cin\'{e}tica n\~{a}o seria finito. \'{E} assim
porque%
\begin{equation}
\psi ^{\ast }\frac{d^{2}\psi }{dx^{2}}=\frac{d}{dx}\left( \psi ^{\ast }\frac{%
d\psi }{dx}\right) -\frac{d\psi ^{\ast }}{dx}\frac{d\psi }{dx},  \label{40}
\end{equation}%
e uma descontinuidade de salto de $\psi $ em $x=0$ faria $d\psi /dx$ ser
proporcional \`{a} fun\c{c}\~{a}o delta de Dirac $\delta \left( x\right) $.
Portanto, a \'{u}ltima parcela do lado direito de (\ref{40}) contribuiria
para o valor esperado da energia cin\'{e}tica com uma parcela proporcional
\`{a}%
\begin{equation}
\lim_{\varepsilon \rightarrow 0}\int_{-\varepsilon }^{+\varepsilon
}dx\;\delta ^{2}\left( x\right) =\infty .
\end{equation}%
Note que a demanda por continuidade da autofun\c{c}\~{a}o em $x=0$ exclui a
possibilidade de uma extens\~{a}o antissim\'{e}trica no caso de $\beta <1$ e
$s=0$.

A conex\~{a}o entre $d\psi /dx$ \`{a} direita e $d\psi /dx$ \`{a} esquerda
da origem pode ser avaliada pela integra\c{c}\~{a}o de (\ref{3}) numa
pequena regi\~{a}o em redor da origem, e para um potencial com o
comportamento ditado por (\ref{pot}) pode ser sumarizada por%
\begin{equation}
\lim_{\varepsilon \rightarrow 0}\left. \frac{d\psi }{dx}\right\vert
_{x=-\varepsilon }^{x=+\varepsilon }=\frac{2m\alpha }{\hbar ^{2}}%
\lim_{\varepsilon \longrightarrow 0}\int_{-\varepsilon }^{+\varepsilon }dx\;%
\frac{\psi }{|x|^{\beta }}.
\end{equation}%
Adotando o valor principal de Cauchy\footnote{%
Para $f\left( x\right) $ singular na origem, pode-se atribuir um sentido
proveitoso \`{a} representa\c{c}\~{a}o integral $\int_{-\infty }^{+\infty
}dx\,\,f\left( x\right) $ por meio da receita que se segue:
\begin{equation}
P\int_{-\infty }^{+\infty }dx\,\,f\left( x\right) =\underset{\delta
\rightarrow 0}{\,\lim }\left( \int_{-\infty }^{-\delta }dx\,\,f\left(
x\right) +\int_{+\delta }^{+\infty }dx\,f\left( x\right) \right) .
\end{equation}%
Tal prescri\c{c}\~{a}o \'{e} conhecida como valor principal de Cauchy de $%
\int_{-\infty }^{+\infty }dx\,\,f\left( x\right) $.} como prescri\c{c}\~{a}o
l\'{\i}cita para atribuir significado \`{a} representa\c{c}\~{a}o integral
cujo integrando \'{e} singular no interior da regi\~{a}o de integra\c{c}\~{a}%
o, pode-se concluir que a derivada primeira de uma autofun\c{c}\~{a}o \'{\i}%
mpar \'{e} sempre cont\'{\i}nua. Entretanto, a autofun\c{c}\~{a}o par requer
aten\c{c}\~{a}o. Nesta \'{u}ltima circunst\^{a}ncia, temos%
\begin{equation}
\left. \frac{d\psi }{dx}\right\vert _{x=0_{+}}-\left. \frac{d\psi }{dx}%
\right\vert _{x=0_{-}}=\left\{
\begin{array}{c}
0, \\
\\
\infty ,%
\end{array}%
\begin{array}{c}
\quad \text{para }\beta =2\text{ e }\alpha >0,\text{ }\quad \text{ou }\beta
<2 \\
\\
\quad \text{para }\beta =2\text{ e }\alpha <0,%
\end{array}%
\right.
\end{equation}%
resultando na exclus\~{a}o das extens\~{o}es sim\'{e}tricas dos casos $1\leq
\beta <2$, e $\beta <1$ com $s=1$.

Um resumo das extens\~{o}es poss\'{\i}veis est\'{a} transcrito na Tabela 2:
\begin{table}[tbp]
\begin{center}
\begin{tabular}{|c||c|}
\hline
&  \\
& extens\~{o}es poss\'{\i}veis \\
&  \\ \hline\hline
&  \\
$\beta =2$ & \text{par e \'{\i}mpar} \\
&  \\ \hline
&  \\
$1\leq \beta <2$ & \'{\i}mpar \\
&  \\ \hline
&  \\
$\beta <1\text{ e }s=0$ & par \\
&  \\ \hline
&  \\
$\beta <1\text{ e }s=1$ & \'{\i}mpar \\
&  \\ \hline
\end{tabular}%
\end{center}
\caption{Poss\'{\i}veis extens\~{o}es sim\'{e}tricas e antissim\'{e}tricas
em fun\c{c}\~{a}o dos par\^{a}metros do potencial.}
\end{table}

A hermiticidade do operador associado com a energia potencial, por causa da
singularidade em $x=0$, depende da exist\^{e}ncia do valor principal de
Cauchy da integral $\int_{-\infty }^{+\infty }dx\,V_{\tilde{n}n}^{\left(
\tilde{p}p\right) }$. Obviamente, o valor principal de Cauchy poderia
consentir um a\-frou\-xa\-men\-to das condi\c{c}\~{o}es de contorno impostas
sobre as autofun\c{c}\~{o}es. Autofun\c{c}\~{o}es mais singulares que essas
anteriormente definidas no semieixo seriam toleradas se na vi\-zi\-nhan\-%
\c{c}a da origem os sinais de $V_{\tilde{n}n}^{\left( \tilde{p}p\right) }$
\`{a} direita e \`{a} esquerda da origem fossem diferentes para quaisquer $p$
e $\tilde{p}$. Por\'{e}m, temos $V_{\tilde{n}n}^{\left( \tilde{p}p\right)
}\left( x<0\right) =\tilde{p}p\,V_{\tilde{n}n}\left( x>0\right) $, de modo
que a integral de $V_{\tilde{n}n}^{\left( \tilde{p}p\right) }$ n\~{a}o seria
finita ao se considerar duas autofun\c{c}\~{o}es com a mesma paridade. Somos
assim conduzidos a preservar a rigidez do crit\'{e}rio de hermiticidade j%
\'{a} estabelecido no problema definido no semieixo.

O crit\'{e}rio de n\~{a}o-degeneresc\^{e}ncia estabelecido na Se\c{c}\~{a}o
3 para o problema definido no semieixo torna-se um fiasco para o problema
definido em todo o eixo no caso em que $\beta =2$. Por causa das condi\c{c}%
\~{o}es de conex\~{a}o entre a autofun\c{c}\~{a}o e sua derivada primeira
\`{a} direita e \`{a} esquerda da origem, e tamb\'{e}m por causa da
hermiticidade do operador associado com a energia potencial, os estados
ligados da equa\c{c}\~{a}o de Schr\"{o}dinger com um potencial que se
comporta como (\ref{pot}) apresentam um espectro degenerado se $\beta =2$, n%
\~{a}o-degenerado e somente com autofun\c{c}\~{o}es \'{\i}mpares se $1\leq
\beta <2$, e n\~{a}o-degenerado com autofun\c{c}\~{o}es pares ($s=0$) e
\'{\i}mpares ($s=1$) se $\beta <1$. O espectro n\~{a}o-degenerado no caso em
que o potencial dominante na origem tem singularidade $1/|x|^{\beta }$ com $%
\beta <1$ \'{e} habitual para sistemas unidimensionais. \'{E} curioso a aus%
\^{e}ncia de autofun\c{c}\~{o}es pares no caso em que o potencial dominante
na origem possui singularidade $1/|x|^{\beta }$ com $1\leq \beta <2$.
Outrossim, \'{e} surpreendente que a degeneresc\^{e}ncia s\'{o} apresente
sua assinatura no caso em que o potencial dominante na origem tem
singularidade $1/x^{2}$.

\section{Dois modelos exemplares}

At\'{e} o momento, temos exclu\'{\i}do classes de potenciais e classes de
solu\c{c}\~{o}es, contudo, as classes de potenciais e classes de solu\c{c}%
\~{o}es que sobejam carecem de modelos espec\'{\i}ficos para a verifica\c{c}%
\~{a}o da concretiza\c{c}\~{a}o de suas possibilidades. Nesta Se\c{c}\~{a}o,
investigaremos as solu\c{c}\~{o}es para $V\left( |x|\right) $ com formas bem
definidas.

Para potenciais com os comportamentos ditados por (\ref{pot}), o
comportamento assint\'{o}tico de $\psi $ expresso por (\ref{asym})
convida-nos a definir $y=2k|x|$ de forma que a autofun\c{c}\~{a}o para todo $%
y\in $ $[0,\infty )$ pode ser escrita como%
\begin{equation}
\psi \left( y\right) =y^{s}\,e^{-y/2}\,w\left( y\right) .  \label{eq4}
\end{equation}%
Por causa do comportamento j\'{a} prescrito para $\psi $, a fun\c{c}\~{a}o $%
w $ deve convergir para uma constante n\~{a}o-nula quando $y\rightarrow 0$,
e n\~{a}o deve crescer mais rapidamente do que $\exp \left(
c_{1}\,y^{c_{2}}\right) $, onde $c_{1}$ \'{e} uma constante arbitr\'{a}ria e
$c_{2}<1$, quando $y\rightarrow \infty $. Isto \'{e} necess\'{a}rio para
garantir que o expoente da exponencial em (\ref{eq4}), viz. $%
-y/2+c_{1}\,y^{c_{2}}$, se comporte como $-y/2$ quando $y\rightarrow \infty $%
.

Veremos mais adiante que as equa\c{c}\~{o}es obedecidas por $w$ para os
problemas pass\'{\i}veis de solu\c{c}\~{o}es anal\'{\i}ticas recaem em equa%
\c{c}\~{o}es diferenciais hipergeom\'{e}tricas confluentes (tamb\'{e}m
chamada de equa\c{c}\~{a}o de Kummer) \cite{abr}
\begin{equation}
y\,\frac{d^{\,2}w\left( y\right) }{dy^{2}}+(b-y)\,\frac{dw\left( y\right) }{%
dy}-a\,w\left( y\right) =0.  \label{kum1}
\end{equation}%
\noindent Buscaremos solu\c{c}\~{o}es particulares de (\ref{kum1}) que
satisfa\c{c}am as condi\c{c}\~{o}es de contorno apropriadas, estabelecidas
no final do par\'{a}grafo anterior. Afortunadamente, veremos tamb\'{e}m que
tais solu\c{c}\~{o}es particulares requerem $b>1$. A solu\c{c}\~{a}o geral
de (\ref{kum1}) \'{e} dada por \cite{abr}
\begin{equation}
w\left( y\right) =C_{1}\,M(a,b,y)+C_{2}\,y^{1-b}\,M(a-b+1,2-b,y),  \label{sg}
\end{equation}%
onde$\ C_{1}$ e $C_{2}$ s\~{a}o constantes arbitr\'{a}rias, e $M(a,b,y)$,
tamb\'{e}m denotada por $_{1}F_{1}\left( a,b,y\right) $, \'{e} a fun\c{c}%
\~{a}o hipergeom\'{e}trica confluente (tamb\'{e}m chamada de fun\c{c}\~{a}o
de Kummer) expressa pela s\'{e}rie \cite{abr}%
\begin{equation}
M(a,b,y)=\frac{\Gamma \left( b\right) }{\Gamma \left( a\right) }%
\sum_{j=0}^{\infty }\frac{\Gamma \left( a+j\right) }{\Gamma \left(
b+j\right) }\,\frac{y^{j}}{j!},  \label{ser}
\end{equation}%
onde $\Gamma \left( z\right) $ \'{e} a fun\c{c}\~{a}o gama. \noindent A fun%
\c{c}\~{a}o gama n\~{a}o tem ra\'{\i}zes e seus polos s\~{a}o dados por $%
z=-n $, onde $n$ \'{e} um inteiro n\~{a}o-negativo \cite{abr}. A fun\c{c}%
\~{a}o de Kummer converge para todo $y$, \'{e} regular na origem ($M(a,b,0)=1
$) e tem o comportamento assint\'{o}tico prescrito por \cite{abr}%
\begin{equation}
M(a,b,y)\,_{\simeq }\,\frac{\Gamma \left( b\right) }{\Gamma \left(
b-a\right) }\,e^{-i\pi a}\,y^{-a}+\frac{\Gamma \left( b\right) }{\Gamma
\left( a\right) }\,e^{y}\,y^{a-b}.  \label{asy}
\end{equation}%
Haja vista que $b>1$, e estamos em busca de solu\c{c}\~{a}o regular na
origem, devemos tomar $C_{2}=0$ em (\ref{sg}). A presen\c{c}a de $e^{y}$ em (%
\ref{asy}) estraga o bom comportamento assint\'{o}tico da autofun\c{c}\~{a}o
j\'{a} ditado por (\ref{asym}). Esta situa\c{c}\~{a}o pode ser remediada
pela considera\c{c}\~{a}o dos polos de $\Gamma \left( a\right) $, e assim
preceituar que um comportamento aceit\'{a}vel para $M(a,b,y)$ ocorre somente
se $a=-n$, com $n\in
\mathbb{N}
$. Neste caso, $M(-n,b,y)$ exibe o comportamento assint\'{o}tico $%
M(-n,b,y)\sim y^{n}$, e a s\'{e}\-rie (\ref{ser}) \'{e} truncada em $j=n$ de
tal forma que o polin\^{o}mio de grau $n$ resultante \'{e} proporcional ao
polin\^{o}mio de Laguerre generalizado $L_{n}^{\left( b-1\right) }\left(
y\right) $ \cite{abr}. \bigskip O polin\^{o}mio de Laguerre generalizado
\'{e} definido pela f\'{o}rmula%
\begin{equation}
L_{n}^{\left( b-1\right) }\left( y\right) =y^{-\left( b+1\right) }e^{y}\frac{%
d^{n}}{dy^{n}}\left( y^{b+n-1}e^{-y}\right) ,\quad b>0,
\end{equation}%
e possui $n$ zeros distintos. Quando $b=1$ o polin\^{o}mio de Laguerre
generalizado \'{e} denotado por $L_{n}\left( y\right) $, e \'{e} chamado
simplesmente de polin\^{o}mio de Laguerre. Haja vista que nosso interesse
recair\'{a} nos casos com $b>1$, podemos assegurar que a autofun\c{c}\~{a}o
do problema ter\'{a} a forma%
\begin{equation}
\psi \left( y\right) =N_{n}\,y^{s}\,e^{-y/2}\,L_{n}^{\left( b-1\right)
}\left( y\right) ,
\end{equation}%
com $n$ nodos no intervalo $\left( 0,\infty \right) $. $N_{n}$ \'{e} uma
constante de normaliza\c{c}\~{a}o.

Vale a pena observar que o comportamento de $w\left( y\right) $ nos extremos
do intervalo, consoante o que foi exposto no final do par\'{a}grafo que
envolve (\ref{eq4}), garante a exist\^{e}ncia de sua transformada de Laplace
(veja, e.g., \cite{but}):%
\begin{equation}
F\left( \sigma \right) =\int_{0}^{\infty }dy\,e^{-\sigma y}w\left( y\right) ,
\label{l1}
\end{equation}%
e assim a solu\c{c}\~{a}o particular de (\ref{kum1}) poderia ter sido obtida
pelo met\'{o}do da transformada de Laplace de $w\left( y\right) $, em concord%
\^{a}ncia com o que foi apregoado em \cite{lap}.

\subsection{$V\left( x\right) =\protect\alpha /|x|^{\protect\beta }$}

\noindent Neste caso, a substitui\c{c}\~{a}o de $\psi \left( y\right) $ dado
por (\ref{eq4}) em (\ref{3}) resulta que $w\left( y\right) $ \'{e} solu\c{c}%
\~{a}o da equa\c{c}\~{a}o
\begin{equation}
y\,\frac{d^{\,2}w\left( y\right) }{dy^{2}}\,+(2s-y)\,\frac{dw\left( y\right)
}{dy}-\left[ s-s\left( s-1\right) y^{-1}+\frac{2m\alpha }{\hbar ^{2}}\left(
2k\right) ^{\beta -2}y^{1-\beta }\right] \,w\left( y\right) =0.  \label{kum}
\end{equation}%
Nota-se que a equa\c{c}\~{a}o (\ref{kum}) se reduz a equa\c{c}\~{o}es
hipergeom\'{e}tricas caso $\beta $ seja igual a $2$ ou $1$:%
\begin{equation}
y\,\frac{d^{\,2}w\left( y\right) }{dy^{2}}+(2s-y)\,\frac{dw\left( y\right) }{%
dy}-s\,w\left( y\right) =0,\quad {\text{para }}\beta =2,
\end{equation}%
e%
\begin{equation}
y\,\frac{d^{\,2}w\left( y\right) }{dy^{2}}+(2-y)\,\frac{dw\left( y\right) }{%
dy}-\left( 1+\frac{m\alpha }{\hbar ^{2}k}\right) \,w\left( y\right) =0,\quad
{\text{para }}\beta =1.
\end{equation}%
Encontramos que solu\c{c}\~{o}es bem comportadas requerem
\begin{equation}
-n=\left\{
\begin{array}{c}
s, \\
\\
1+\frac{m\alpha }{\hbar ^{2}k},%
\end{array}%
\begin{array}{c}
{\text{para }}\beta =2 \\
\\
\quad {\text{para }}\beta =1.%
\end{array}%
\right.  \label{N1}
\end{equation}%
Da\'{\i} podemos concluir que n\~{a}o h\'{a} solu\c{c}\~{a}o se $\beta =2$,
pois $s>1/2$. Entretanto, \ se $\beta =1$ h\'{a} solu\c{c}\~{o}es somente no
caso em que $\alpha <0$, como esperado. Tais solu\c{c}\~{o}es, com $b=2$, s%
\~{a}o expressas em termos de $L_{n}^{\left( 1\right) }\left( y\right) $ por:%
\begin{eqnarray}
E_{n} &=&-\frac{m\alpha ^{2}}{2\hbar ^{2}\left( n+1\right) ^{2}}  \notag \\
&&  \label{S1} \\
\psi _{n}\left( |x|\right) &=&N_{n}\,|x|\,\exp \left( -\frac{m|\alpha |}{%
\hbar ^{2}\left( n+1\right) }|x|\right) \,L_{n}^{\left( 1\right) }\left(
\frac{2m|\alpha |}{\hbar ^{2}\left( n+1\right) }|x|\right).  \notag
\end{eqnarray}%
Para $V\left( x\right) =-|\alpha |/|x|$, encontramos um conjunto infinito de
estados ligados. N\~{a}o h\'{a} limite inferior imposto sobre $\alpha $,
basta que $\alpha $ seja negativo.

\subsection{$V\left( x\right) =\protect\alpha _{1}/|x|+\protect\alpha %
_{2}/x^{2},\quad \protect\alpha _{1}\neq 0$}

Neste caso,%
\begin{equation}
y\,\frac{d^{\,2}w\left( y\right) }{dy^{2}}+(2s-y)\,\frac{dw\left( y\right) }{%
dy}-\left( s+\frac{m\alpha _{1}}{\hbar ^{2}k}\right) \,w\left( y\right) =0,
\end{equation}%
e as solu\c{c}\~{o}es bem comportadas demandam
\begin{equation}
-n=s+\frac{m\alpha _{1}}{\hbar ^{2}k}.  \label{N2}
\end{equation}%
Necessariamente com $\alpha _{1}<0$, as solu\c{c}\~{o}es com $s>1/2$ ($b>1$)
s\~{a}o expressas em termos de $L_{n}^{\left( 2s-1\right) }\left( y\right) $
por:%
\begin{eqnarray}
E_{n} &=&-\frac{m\alpha _{1}^{2}}{2\hbar ^{2}\left( n+s\right) ^{2}}  \notag
\\
&&  \label{S2} \\
\psi _{n}\left( |x|\right) &=&N_{n}\,|x|^{s}\,\exp \left( -\frac{m|\alpha
_{1}|}{\hbar ^{2}\left( n+s\right) }|x|\right) \,L_{n}^{\left( 2s-1\right)
}\left( \frac{2m|\alpha _{1}|}{\hbar ^{2}\left( n+s\right) }|x|\right) .
\notag
\end{eqnarray}%
Aqui,
\begin{equation}
s=\frac{1}{2}\left( 1+\sqrt{1+\frac{8m\alpha _{2}}{\hbar ^{2}}}\right) ,%
\text{ }\quad \alpha _{2}>\alpha _{c}.
\end{equation}%
Para $V\left( x\right) =-|\alpha _{1}|/|x|+\alpha _{2}/x^{2}$, com $\alpha
_{1}\neq 0$, encontramos um conjunto infinito de estados ligados com nenhum
limite inferior imposto sobre $\alpha _{1}$, basta que $\alpha _{1}<0$. Para
$\alpha _{2}=0$, os resultados coincidem com aqueles encontrados na subse%
\c{c}\~{a}o anterior, como deveria acontecer. Qualquer que seja $\alpha
_{2}>\alpha _{c}$, as solu\c{c}\~{o}es s\~{a}o fisicamente aceit\'{a}veis,
ainda que no intervalo$\ \alpha _{c}<\alpha _{2}<0$ as autofun\c{c}\~{o}es
possuam derivadas primeiras singulares na origem. \'{E} mesmo assim,
contanto que o par\^{a}metro $\alpha _{2}$ seja maior que $\alpha _{c}$, o
par caracter\'{\i}stico $\left( E_{n},\psi _{n}\right) $ constitui uma solu%
\c{c}\~{a}o permiss\'{\i}vel do problema proposto. Visto como fun\c{c}\~{a}o
de $\alpha _{2}$, ao passar por $\alpha _{2}=0$, a forma do potencial sofre
uma mudan\c{c}a dr\'{a}stica: \ passa de um po\c{c}o sem fundo com
singularidade negativa na origem quando $\alpha _{c}<\alpha _{2}\leq 0$ para
um po\c{c}o com fundo com singularidade positiva na origem quando $\alpha
_{2}>0$. As autofun\c{c}\~{o}es sempre satisfazem \`{a} condi\c{c}\~{a}o
homog\^{e}nea de Dirichlet na origem. As derivadas primeiras das autofun\c{c}%
\~{o}es \ s\~{a}o infinitas na origem se $\alpha _{c}<\alpha _{2}<0$. Para $%
\alpha _{2}\geq 0$, contudo, as derivadas primeiras s\~{a}o finitas, sendo
nulas caso $\alpha _{2}>0$. \'{E} instrutivo observar que esta transi\c{c}%
\~{a}o de fase n\~{a}o se manifesta no espectro.

Nas Figuras 1 e 2
\begin{figure}[th]
\begin{center}
\includegraphics[width=10cm]{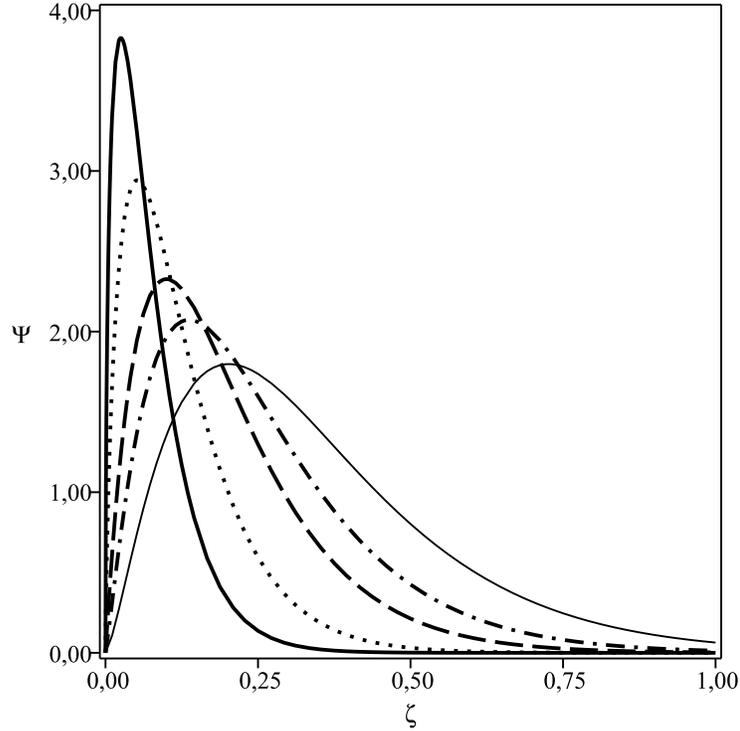} \label{fig:Fig1}
\end{center}
\par
\vspace*{-0.1cm}
\caption{Autofun\c{c}\~{a}o normalizada do estado fundamental definido no
semieixo em fun\c{c}\~{a}o de $\protect\zeta =|x|/\protect\lambda $, para $%
\protect\alpha _{1}/(\hbar c)=-10$. As linhas cont\'{\i}nua espessa,
pontilhada, tracejada, ponto-tracejada e cont\'{\i}nua delgada para os casos
com $m\protect\alpha _{2}/\hbar ^{2}$ igual a $-0,124999$, $-0,1$, $0$, $%
+0,1 $ e $0,3$, respectivamente.}
\end{figure}
\begin{figure}[th]
\begin{center}
\includegraphics[width=10cm]{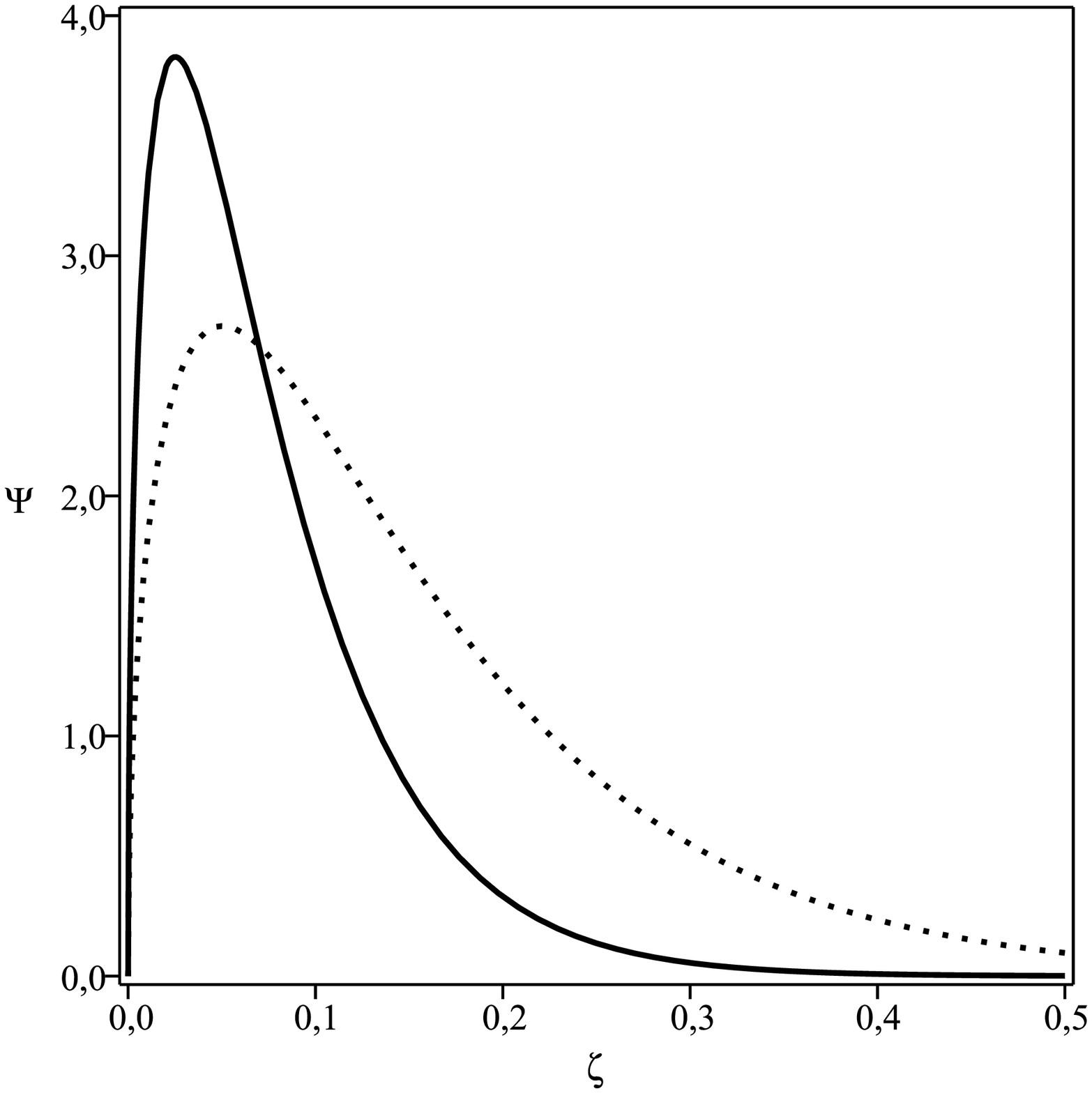} \label{fig:Fig2}
\end{center}
\par
\vspace*{-0.1cm}
\caption{Autofun\c{c}\~{a}o normalizada do estado fundamental definido no
semieixo em fun\c{c}\~{a}o de $\protect\zeta =|x|/\protect\lambda $, para $m%
\protect\alpha _{2}/\hbar ^{2}=-0,124999$. As linhas cont\'{\i}nua e
pontilhada para os casos com $\protect\alpha _{1}/(\hbar c)$ igual a $-10$ e
$-5$, respectivamente.}
\end{figure}
ilustramos o comportamento da autofun\c{c}\~{a}o ($\Psi =\sqrt{\lambda }%
\;\psi $) para o estado fundamental em fun\c{c}\~{a}o de $\zeta =|x|/\lambda
$, onde
\begin{equation}
\lambda =\frac{\hbar }{mc}
\end{equation}%
\'{e} o comprimento de onda Compton da part\'{\i}cula ($c$ \'{e} a
velocidade da luz). A nor\-ma\-li\-za\-\c{c}\~{a}o foi realizada por m\'{e}%
todos num\'{e}ricos mas poderia ter sido obtida por meio de f\'{o}rmulas
envolvendo os polin\^{o}mios de Laguerre generalizados constantes na Ref.
\cite{abr}. Na Figura 1, fixamos $\alpha _{1}$ e consideramos cinco valores
ilustrativos de $\alpha _{2}$. Na Figura 2, fixamos $\alpha _{2}$ na vizinhan%
\c{c}a do valor cr\'{\i}tico $\alpha _{c}$ e consideramos dois valores
ilustrativos de $\alpha _{1}$.

A compara\c{c}\~{a}o entre as cinco curvas da Figura 1 mostra que a part%
\'{\i}cula tende a evitar a origem mais e mais \`{a} medida que $\alpha _{2}$
aumenta. A part\'{\i}cula nunca colapsa para o ponto $x=0$ (em contradi\c{c}%
\~{a}o com a afirma\c{c}\~{a}o patente na Ref. \cite{bg}), e certamente h%
\'{a} um estado fundamental com energia igual a $E_{0}=-2m\alpha
_{1}^{2}/\left( 2\hbar s\right) ^{2}$.

A compara\c{c}\~{a}o entre as duas curvas da Figura 2, para $\alpha
_{2}\gtrsim \alpha _{c}$, mostra que a part\'{\i}cula tende a evitar a
origem mais e mais \`{a} medida que $\alpha _{1}$ aumenta.

No caso das extens\~{o}es para todo o eixo, a forma do potencial $V\left(
x\right) =-|\alpha _{1}|/|x|+\alpha _{2}/x^{2}\ $(com $\alpha _{1}\neq 0$),
ao passar por $\alpha _{2}=0$, troca de um po\c{c}o sem fundo com
singularidade negativa em $x=0$ quando $\alpha _{c}<\alpha _{2}\leq 0$ para
um po\c{c}o duplo com singularidade positiva em $x=0$ \ quando $\alpha
_{2}>0 $. Autofun\c{c}\~{o}es possuem derivadas primeiras finitas na origem
se $\alpha _{2}\geq 0$. Para $\alpha _{c}<\alpha _{2}<0$, contudo, as
derivadas primeiras s\~{a}o infinitas na origem. Esta transi\c{c}\~{a}o de
fase em $\alpha _{2}=0$ tamb\'{e}m se manifesta no grau de degeneresc\^{e}%
ncia do espectro.

\section{Coment\'{a}rios finais}

Dissertamos sobre o problema de estados ligados no \^{a}mbito da equa\c{c}%
\~{a}o de Schr\"{o}dinger unidimensional para uma classe de potenciais $%
V\left( |x|\right) $ que se anulam no infinito e a\-pre\-sen\-tam
singularidade dominante na origem na forma $\alpha /|x|^{\beta }$ ($0<\beta
\leq 2$) por meio dos comportamentos das autofun\c{c}\~{o}es na vizinhan\c{c}%
a da origem e seus comportamentos assint\'{o}ticos. Armados com o crit\'{e}%
rio de hermiticidade dos operadores associados com quantidades f\'{\i}sicas
observ\'{a}veis, e com o valor principal de Cauchy para atribuir significado
\`{a} representa\c{c}\~{a}o integral cujo integrando \'{e} singular no
interior da regi\~{a}o de integra\c{c}\~{a}o, conclu\'{\i}mos que:

\begin{itemize}
\item o espectro \'{e} negativo;

\item n\~{a}o h\'{a} estados ligados se o potencial com comportamento
dominante na origem do tipo inversamente quadr\'{a}tico for fortemente
atrativo ($\alpha /x^{2}$ com $\alpha \leq \alpha _{c}$);

\item o espectro do problema definido no semieixo \'{e} n\~{a}o-degenerado;

\item o espectro do problema definido em todo o eixo \'{e} n\~{a}%
o-degenerado se $\beta <2$, e duplamente degenerado se $\beta =2$;

\item somente autofun\c{c}\~{o}es antissim\'{e}tricas ocorrem no caso $1\leq
\beta <2$;

\item autofun\c{c}\~{o}es sim\'{e}tricas e antissim\'{e}tricas ocorrem no
caso $\beta <1$.
\end{itemize}

Em seguida, tratamos de modelos exatamente sol\'{u}veis. Consideramos
potenciais da forma $V\left( |x|\right) =\alpha _{1}/|x|+\alpha _{2}/x^{2}$,
e recorrendo ao conhecimento de solu\c{c}\~{o}es de equa\c{c}\~{o}es
diferenciais hipergeom\'{e}tricas confluentes, demonstramos que inexistem
solu\c{c}\~{o}es no caso em que $\alpha _{1}=0$, e que h\'{a} um n\'{u}mero
infinito de estados ligados no caso em que $\alpha _{1}<0$ e $\alpha
_{2}>-\hbar ^{2}/\left( 8m\right) $. Em particular, reafirmamos as conclus%
\~{o}es obtidas na Ref. \cite{xia} referentes ao \'{a}tomo de hidrog\^{e}nio
unidimensional definido em todo o eixo: todo o espectro tem energia finita e
autofun\c{c}\~{o}es antissim\'{e}tricas.

Nossos resultados para o problema definido no semieixo podem ser
generalizados para o problema tridimensional por meio do acr\'{e}scimo de $%
l\left( l+1\right) \hbar ^{2}/\left( 2mx^{2}\right) $ ao potencial e pela
substitui\c{c}\~{a}o de $\psi \left( |x|\right) $ por $|x|R\left( |x|\right)
$, onde $l$ \'{e} o n\'{u}mero qu\^{a}ntico orbital e $R\left( |x|\right) $
\'{e} a fun\c{c}\~{a}o radial. Desta forma, $\alpha $ deve ser substitu\'{\i}%
do por $\alpha +l\left( l+1\right) \hbar ^{2}/\left( 2m\right) $ se $\beta
=2 $, e $\alpha _{2}$ por $\alpha _{2}+l\left( l+1\right) \hbar ^{2}/\left(
2m\right) $ na Sec. 5.1. Entretanto, deve-se observar que a parcela $l\left(
l+1\right) \hbar ^{2}/\left( 2mx^{2}\right) $ representar\'{a} a
singularidade dominante na origem se $\beta <2$ e $l\neq 0$, em vez de $%
\alpha /|x|^{\beta }$. Al\'{e}m disto, a solu\c{c}\~{a}o que n\~{a}o
satisfaz \`{a} condi\c{c}\~{a}o de Dirichlet homog\^{e}nea na origem, essa
com $s=0$ para $\beta <1$, torna-se uma solu\c{c}\~{a}o esp\'{u}ria para o
caso tridimensional. Isto se d\'{a} porque a exist\^{e}ncia de uma fun\c{c}%
\~{a}o radial com o comportamento na vizinhan\c{c}a da origem ditado por $%
1/|x|$ requeriria a presen\c{c}a de uma fun\c{c}\~{a}o delta de Dirac no
potencial \cite{sha}.

\vspace{15cm}

\noindent{\textbf{Agradecimentos}}

O autores s\~{a}o gratos \`{a} CAPES, ao CNPq e \`{a} FAPESP pelo apoio
financeiro.


\newpage

\begin{thebibliography}{99}
\bibitem{cas} K.M. Case, Phys. Rev. \textbf{80}, 797 (1950); F.L. Scarf,
Phys. Rev. \textbf{109}, 2170 (1958); A. Pais e T.T. Wu, Phys. Rev. \textbf{%
134}, B1303 (1964); E. Ferreira, J. Sesma e P.L. Torres, Prog. Theor. Phys.
\textbf{43}, 1 (1970); W.M. Frank, D.J. Land e R.M. Spector, Rev. Mod. Phys.
\textbf{43}, 36 (1971); H.G. Miller, J. Math. Phys. \textbf{35}, 2229
(1994); V.C. Aguilera-Navarro e A.L. Coelho, Phys. Rev. A \textbf{49}, 1477
(1994); G. Esposito, J. Phys. A \textbf{31}, 9493 (1998); A. Das e S.A.
Pernice, Nucl. Phys. B \textbf{561}, 357 (1999); A.M. Essin e J.D.
Griffiths, Am. J. Phys. \textbf{74}, 109 (2006).

\bibitem{rbef} D.R.M. Pimentel e A.S. de Castro, arXiv:1304.0492 (a ser
publicado na Revista Brasileira de Ensino de F\'{\i}sica).

\bibitem{xia} D. Xianxi, J. Dai e J. Dai, Phys. Rev. A \textbf{55}, 2617
(1997).

\bibitem{yep} H.N. N\'{u}\~{n}ez-Y\'{e}pes, A.L. Salas-Brito e D.A. Solis,
Phys. Rev. A \textbf{83}, 064101 (2011).

\bibitem{spe} H.N. Spector e J. Lee, Am. J. Phys. \textbf{53}, 248 (1985);
R.E. Moss, Am. J. Phys. \textbf{55}, 397 (1987); C.-L. Ho e V.R. Khalilov,
Phys. Rev. D \textbf{63}, 027701 (2001).

\bibitem{asc1} A.S. de Castro, Phys. Lett. A \textbf{338}, 81 (2005); A.S.
de Castro, Ann. Phys. (N.Y.) \textbf{316}, 414 (2005).

\bibitem{dom} N. Dombey, Phys. Lett. A \textbf{360}, 62 (2006); A.S. de
Castro, Phys. Lett. A \textbf{369}, 380 (2007).

\bibitem{bar} G. Barton, J. Phys. A \textbf{40}, 1011 (2007).

\bibitem{kra} A. Kratzer, Z. Phys. \textbf{3}, 289 (1920); G. Van Hooydonk,
J. Mol. Struc. Theochem \textbf{109}, 87 (1984); G. Van Hooydonk, Eur. J.
Inorg. Chem. \textbf{1999}, 1617 (1999).

\bibitem{ulrich} H. Ulrich e P. Hess, Chem. Phys. \textbf{47}, 481 (1980).

\bibitem{bg} V.G. Bagrov e D.M. Gitman, \textit{Exact Solutions of
Relativistic Wave Equations} (Kluwer, Dordrecht, 1990).

\bibitem{lan} L.D. Landau e E.M. Lifshitz, \textit{Quantum Mechanics}
(Pergamon, Nova Iorque, 1958); I.I. Gol'dman e V.D. Krivchenkov, \textit{%
Problems in Quantum Mechanics} (Pergamon, Londres, 1961); F. Constantinescu
e E. Magyari, \textit{Problems in Quantum Mechanics} (Pergamon, Oxford,
1971); D. ter Haar, \textit{Problems in Quantum Mechanics} (Pion, Londres,
1975); S. Fl\"{u}gge, \textit{Practical Quantum Mechanics} (Springer-Verlag,
Berlim, 1999).

\bibitem{asc2} A.S. de Castro, Phys. Lett. A \textbf{328}, 289 (2004); L.K.
Sharma e J. Fiase, Chin. Phys. Lett. \textbf{21}, 1893 (2004); A.S. de
Castro, Int. J. Mod. Phys. A \textbf{21}, 2321 (2006); A.S. de Castro, J.
Math. Phys. \textbf{51}, 102302 (2010).

\bibitem{but} E. Butkov, \textit{F\'{\i}sica Matem\'{a}tica} (LTC, Rio de
Janeiro, 1988).

\bibitem{abr} M. Abramowitz e I.A. Stegun, \textit{Handbook of Mathematical
Functions} (Dover, Toronto, 1965).

\bibitem{lap} D.R.M. Pimentel e A.S. de Castro, Eur. J. Phys. \textbf{34},
199 (2013).

\bibitem{sha} R. Shankar, \textit{Principles of Quantum Mechanics} (Plenum
Press, Nova Iorque, 1994).
\end{thebibliography}
\end{document}